\documentclass[review, 1p]{elsarticle}

\usepackage{amsmath,amssymb,bm,booktabs,color}
\usepackage{multirow}
\usepackage{lineno}
\usepackage[T1]{fontenc} % if needed
\usepackage[utf8]{inputenc}
\usepackage{graphicx}% Include figure files
\usepackage{dcolumn}% Align table columns on decimal point
\usepackage{bm}% bold math
\usepackage{float}
\usepackage{multirow}
\usepackage{verbatim}
\usepackage{amsmath}
\usepackage{graphicx,subfigure,epsfig}
\usepackage{mathrsfs}
\usepackage{hyperref}
\hypersetup{
    colorlinks=true,
    citecolor=blue,
    linkcolor=blue,
    filecolor=magenta,
    urlcolor=black,}%cyan lianjieyanse
\usepackage{color}
\usepackage{xcolor}
\newcommand{\be}{\begin{equation}}
\newcommand{\ee}{\end{equation}}

\biboptions{sort&compress}

\journal{the European Physical Journal C}

\begin{document}

\begin{frontmatter}

\title{Probing black holes in a dark matter spike of M87 using quasinormal modes}

%%Group authors per affiliation: \corref{mycorrespondingauthor}
\author[GZU]{Dong Liu}
\ead{dongliuvv@yeah.net}

\author[GZUFE]{Yi Yang }
\ead{yiyang@mail.gufe.edu.cn}

\author[GZU]{Zheng-Wen Long \corref{mycorrespondingauthor}}
\cortext[mycorrespondingauthor]{Corresponding author}
\ead{zwlong@gzu.edu.cn}

\address[GZU]{College of Physics, Guizhou University, Guiyang, 550025, China}
\address[GZUFE]{School of Mathematics and Statistics, Guizhou University of Finance and Economics, Guiyang, 550025, China}

\begin{abstract}
\indent Dark matter density can be significantly enhanced by the supermassive black hole at the galactic center, leading to a structure called dark matter spike. Dark matter spike may change the spacetime properties of black holes that constitute deviations from GR black holes. Based on these interesting background, we construct a set of solutions of black holes in a dark matter spike under the Newtonian approximation and full relativity. Combining the mass model of M87, we study the quasinormal modes of  black holes in the scalar field and axial gravitational perturbation, then compared them with Schwarzschild black hole. Besides, the impacts of dark matter on the quasinormal mode of black holes have been studied in depth. In particular, in the axial gravitational perturbation, our results show that the impacts of dark matter spike on the quasinormal mode of black holes can reach up to $10^{-4}$. These new features from quasinormal mode of black holes under the Newtonian approximation and full relativity may provide some help for the establishment of the final dark matter model, and provide a new thought for the indirect detection of dark matter.
\end{abstract}


\end{frontmatter}

%\linenumbers

\section{\label{sec1} Introduction}
Black holes are the beautiful but mysterious celestial bodies in our universe. In General Relativity (GR), an isolated black hole in equilibrium can usually be described simply by three parameters, such as mass, angular momentum, and charge. However, such an isolated black hole hardly exists in the real universe. There is always the complex distribution of matter near the black hole in the center of the galaxy, such as the core of the galaxy, accretion disk, dark matter or other planets. Therefore, a black hole will always interact with its surrounding environment, causing the black hole to be in a state of perturbation. Now, the LIGO/Virgo collaboration has detected the gravitational wave (GW) emitted by the merger of binary black holes, and the GW may go through three stages: inspired, merger, and ringdown \cite{LIGO1, LIGO2, LIGO3, LIGO4}. Through gravitational wave, we may be able to understand the true structure of black holes or the matter near black holes. For the ringdown stage of gravitational wave, it can usually be described by the superposition of complex frequency damping exponents, which we call quasinormal mode (QNM) \cite{Konoplya:2011qq}. The real part of the QNM represents the oscillation frequency, and the imaginary part represents the decay rate. The QNM is a special fingerprint of a black hole, which can be used as a tool to detect GR. Therefore, GW detectors may detect some new physical properties in the future, such as GW echoes \cite{Vicente:2018mxl, Bueno:2017hyj, Maselli:2017cmm, Cardoso:2016oxy, Cardoso:2016rao,Dong:2020odp,Yang:2021cvh,Rahman:2021kwb,Ghosh:2022mka}. 

On the other hand, more and more observational evidence indirectly shows the existence of dark matter, such as the rotation curve of galaxies, the large-scale structure of the universe and the cosmic microwave background radiation. Based on these observational facts, astronomers have proposed many dark matter models used to study various types of dark matter, such as cold dark matter (CDM) model, warm dark matter (WDM) model, scalar field dark matter (SFDM) model, and self-interaction dark matter (SIDM) model. Some interesting achievements in the scalar field dark matter model are reviewed by T. Matos et al \cite{Matos:2023usa}. For cold dark matter particles, F. Navarro et al. used N-body simulation to give the cold dark matter distribution of the dark matter halo, that is the NFW profile \cite{Navarro:1995iw}. T. Matos et al. derived the space-time geometry associated with NFW profiles based on Newtonian theory and relativistic analysis \cite{Matos:2003nb}. Y. Yang et al. studied a black hole surrounded by a dark matter halo \cite{Yang:2023tip, Liu:2023oab}. Z. Shen et al. gave five analytical models of black holes immersed in dark matter halo in their recent work \cite{Shen:2023erj}. R.A. Konoplya et al. obtained the exact solutions for the metric immersed in the different dark matter halos from Einstein field equations \cite{Konoplya:2021ube,Konoplya:2022hbl}. On the other hand, P. Gondolo et al. showed that if there are cold dark matter particles near the black hole, the cold dark matter will be accreted into a spike structure by the strong gravity of the black hole based on Newtonian theory and relativistic analysis \cite{Gondolo:1999ef}. Based on full relativistic analysis, L. Sadeghian et al. gave the modified density profile of dark matter spike \cite{Sadeghian:2013laa}. About the new relevant studies between dark matter spikes and black holes, we list as follows. F. Ferrer et al. used Kerr-like metric immersed in a dark matter spike to study the impacts of black hole spin on the distribution of dark matter \cite{Ferrer:2017xwm}. Z. Xu et al. gave the space-time geometry of the black hole under a dark matter spike based on a relativistic analysis, and then extended it to the case of rotation \cite{Xu:2021dkv}. S. Nampalliwar et al. derive the space-time geometry of the Kerr-like black hole at the center of Sgr A in a dark matter spike and study the black hole's shadow \cite{Nampalliwar:2021tyz}. G. Li et al. showed the possibility of detecting dark matter spikes via gravitational waves from extreme-mass-ratio spirals \cite{Li:2021pxf}. N. Speeney and others studied the impact of relativistic corrections on the detectability of gravitational wave with a dark matter spike \cite{Speeney:2022ryg}. These new researches may have provided some effective methods for indirect exploration of dark matter.

In fact, quasinormal mode, as an observable phenomenon, can be used to study and analyze black holes and the distribution of dark matter spike. Cardoso et al. develop a general, fully relativistic formalism for the study of gravitational wave emission in non-vacuum astrophysical environments \cite{Cardoso:2022whc}. D. Liu et al. used the QNM to study black holes in a dark matter halo \cite{Liu:2021xfb,Liu:2022ygf}. C. Zhang et al. studied the QNM of black holes in a dark matter halo and then studied the impacts of dark matter parameters on the QNM of black holes \cite{Zhang:2021bdr,Zhang:2022roh}. R. Daghigh et al. used scalar field perturbation to study the gravitational ringdown of black holes in M87 and Sgr A \cite{Daghigh:2022pcr}. Y. Zhao et al. used axial gravitational perturbation to study the QNM of black holes in in Milky Way \cite{Zhao:2023tyo}. Therefore, in this work, combining the mass model of M87, we constructed a set of solutions of black holes in a dark matter spike under the Newtonian approximation and full relativity. We hope to be able to predict the types of black holes at the galactic center of M87 through the QNM, thereby know which one between the Newtonian approximation and full relativity is closer to the observational facts. Besides, we are also interested in the impacts of dark matter spike on QNM of black holes. These new results based on the QNM may provide more possibilities for local measurement of dark matter. Finally, we study and explore the possibility of space-based detection detecting the impacts of dark matter on the QNM of black holes. 

This paper is organized as follows. In Sect. \ref{s2}, we introduce the density profile of dark matter near supermassive black holes. In Sect. \ref{s3}, from Einstein field equations, we construct a set of black hole solutions based on Newtonian approximation and full relativity. In Sect. \ref{s4}, we obtained the master equation of black holes in a dark matter spike under the scalar field and axial gravitational perturbations. In Sect. \ref{s5}, we introduce some numerical methods for calculating quasinormal mode, namely the time domain integration method, the Prony method and the WKB method. In Sect. \ref{s6}, we study the quasinormal mode of the black hole in a dark matter spike of M87 and compare it with the Schwarzschild black hole. The impacts of dark matter spike on the quasinormal mode of black holes have been studied. Finally, we explore the possibility of space-based detectors detecting the impacts of dark matter parameters on the quasinormal mode of black holes. The Sect. \ref{s7} is our summary. In this work, the Greek indices is from 1 to 4, and we use natural unit of $G=c=2 M_\text{BH}=1$, the radius of the Schwarzschild black hole is given by $r_\text{s}=2 GM_\text{BH}/c^2$.

\section{Density profile of a dark matter spike near the black hole}\label{s2}
\begin{figure*}[t!]
\centering
{
\label{fig:b}
\includegraphics[width=0.48\columnwidth]{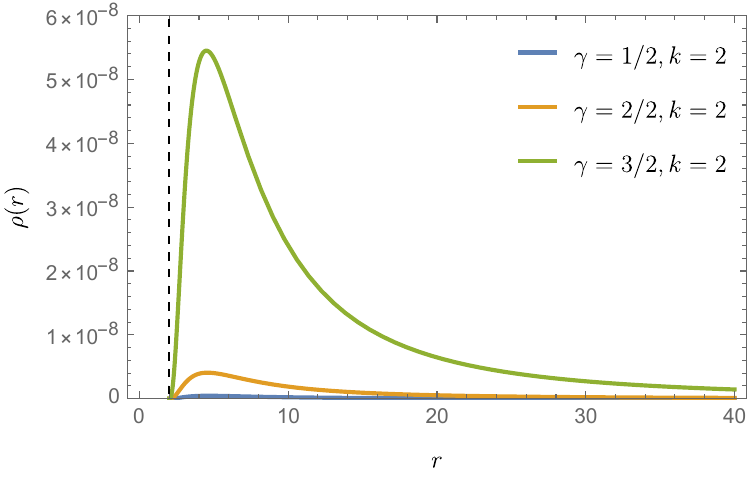}
}
{
\label{fig:b}
\includegraphics[width=0.48\columnwidth]{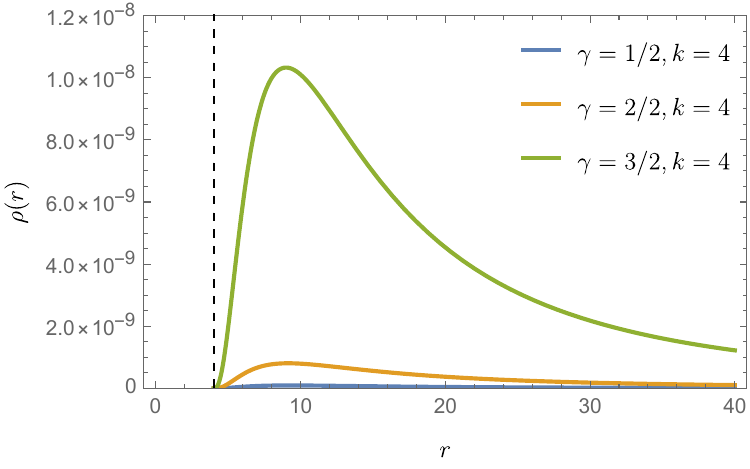}
}
\caption{Density profile of DM spike as a function of variate $r$ under the Newtonian approximation ($k=4$) and full relativity ($k=2$) in the galactic center of M$87$. The dotted lines in the panels represent $r=2R_\mathrm{s}, r=4R_\mathrm{s}$, respectively. When the DM density is smaller than them, it indicates that dark matter disappears or falls into a black hole. The parameters we used are $M_\text{BH}=1/2, \alpha_\gamma=0.1, \rho_0=6.9 \times 10^6 M_\odot/\text{kpc}^3, r_0=91.2\text{kpc}$.}
\label{f1}
\end{figure*}

Due to the strong gravity of the black hole itself, the structure of the dark matter (DM) halo near the black hole is usually considered to be a ``spike''. The density profile of this DM spike can be well described by two parameters, namely the halo density and characteristic radius \cite{Sadeghian:2013laa,Gondolo:1999ef}
\begin{equation}
\rho_\text{spike}(r)=\rho_\text{sp}\left(1- k R_\text{s} / r\right)^{3}\left( r_{\text{sp}} / r\right)^{\gamma _\text{sp} },
\label{e1}
\end{equation}
where, $k$ is zero point parameter, $\rho_\text{sp}$ is the density of DM spike, $r_\text{sp}$ is the radius of DM spike, $R_\text{s}$ is the radius of Schwarzschild black hole, and $\gamma _\text{sp}$ is spike slope,
\begin{equation}
\begin{aligned}
r_\text{sp} = \alpha _\gamma r_0(M_{\text{BH}}/\rho_0r_0^3)^{1/(3-\gamma )}, \quad \rho_ \text{sp} = \rho_0(r_\text{sp}/r_0)^{-\gamma }, \quad \gamma _\text{sp}=(9-2\gamma )/(4-\gamma ),
\label{e2}
\end{aligned}
\end{equation}
here, $\rho_0$, $r_0$ are the halo density and characteristic radius in the galactic center, respectively, $M_\text{BH}$ is the mass of the black hole, $\gamma$ is the power-law index and $\alpha_\gamma$ is normalization scale factor. Here, we adopt the power-law index $\gamma$ from Refs. \cite{Gondolo:1999ef, McMillan:2016jtx, Shen:2023kkm} and mainly focus on the case of $\gamma \in [1/2,3/2]
$ and then $\gamma_\text{sp} \in [16/7,12/5]$. For Eq. (\ref{e1}), the zero point parameters $k=4$ and $k=2$ simply correspond to the case of the Newtonian approximation \cite{Gondolo:1999ef} and full relativity \cite{Sadeghian:2013laa,Tang:2020jhx}, respectively. For the convenience of discussion, the parameters related to the dark matter need to be converted in the black hole units (BHU) by the following formula: $r_\text{0}(\text{BHU})=r_\text{0} / (2 G M_{\text{BH}}/c^2) \times r_\text{BH}$ and $\rho_\text{0}(\text{BHU})=\rho_\text{0} /(M_\text{BH}/(4/3 \pi (2 G M_{\text{BH}}/c^2)^3))\times \rho_{\text{BH}}$. Here, we are interested in the impacts of dark matter at the galactic center of M87. The relevant parameters of M87 are $\rho_0=6.9 \times 10^6 M_\odot/\text{kpc}^3, r_0=91.2 \text{kpc}$ and the BH mass at the galactic center is $M=6.5 \times 10^9 M_\odot$ \cite{EventHorizonTelescope:2019dse}. Therefore, in the black hole units, we have the parameters\footnote{ Unless otherwise stated, our calculations in this work use the density profile in a DM spike and the BH mass model of M87.}, $\rho_0 (\text{BHU})  \approx 1.41\times10^{-22}$, $r_0 (\text{BHU}) \approx 1.42 \times 10^8$. In Figure \ref{f1}, we show these two density profiles of dark matter spike as a function of variate $r$ in the galactic center of M$87$. By comparing these figures, our results show that under the Newtonian approximation ($k=4$) and the full relativistic case ($k=2$), the density profile of dark matter spike increases with increasing of the power-law index $\gamma$. Besides, under the fully relativistic case, the density profile of dark matter spike is obviously enhanced, which means that the impact of dark matter spike on the black hole may be increased.
%%%%%%%%%%%%%%%%%%%%%%%%
%%%%%%%%%%%%%%%%%%%%%%%%

Based on the density profile of the dark matter spike in Eq. (\ref{e1}), we can obtain the mass distribution of the dark matter spike
\begin{equation}
\begin{aligned}
M_{\text{spike}}(r)&=4\pi \int_{0}^{r}\rho_{\text{spike}} (r')r'^2dr' = \frac{4k^3\pi R_\text{s}^3 (r/r_{\text{sp}})^{-\gamma _\text{sp} }\rho _\text{sp}}{\gamma _\text{sp} } - \frac{12k^2\pi r R_\text{s}^2 (r/r_{\text{sp}})^{-\gamma _\text{sp} }\rho _\text{sp}}{\gamma _\text{sp}-1 }\\
& + \frac{12k \pi r^2 R_\text{s} (r/r_{\text{sp}})^{-\gamma _\text{sp} }\rho _\text{sp}}{\gamma _\text{sp}-2 } - \frac{4\pi r^3 (r/r_{\text{sp}})^{-\gamma _\text{sp} }\rho _\text{sp}}{\gamma _\text{sp} -3}\\ 
&-\frac{24k^3 \pi R_\text{s}^3 (k R_\text{s}/r_{\text{sp}})^{-\gamma _\text{sp} } \rho _\text{sp}}{\gamma _\text{sp} (\gamma _\text{sp} -1)(\gamma _\text{sp} -2)(\gamma _\text{sp}-3)}.
\label{e3}
\end{aligned}
\end{equation}
%%%%%%%%%%%%%%%%
In the Newtonian theory, for the test particles at the equatorial plane of static spherical symmetric spacetime, their tangential velocity can be usually determined by the mass distribution of dark matter spike \cite{Matos:2003nb}. Therefore, the tangential velocity $V$ of the test particles can be defined as 
\begin{equation}
\begin{aligned}
V_{\text{spike}}(r)&=\sqrt{M_\text{spike}/r}=\left[\left(\frac{4k^3\pi R_\text{s}^3 (r/r_{\text{sp}})^{-\gamma _\text{sp} }\rho _\text{sp}}{\gamma _\text{sp} } - \frac{12k^2\pi r R_\text{s}^2 (r/r_{\text{sp}})^{-\gamma _\text{sp} }\rho _\text{sp}}{\gamma _\text{sp}-1 }\right.\right. \\
&+ \frac{12k \pi r^2 R_\text{s} (r/r_{\text{sp}})^{-\gamma _\text{sp} }\rho _\text{sp}}{\gamma _\text{sp}-2 } - \frac{4\pi r^3 (r/r_{\text{sp}})^{-\gamma _\text{sp} }\rho _\text{sp}}{\gamma _\text{sp} -3}\\
&\left.\left.-\frac{24k^3 \pi R_\text{s}^3 (k R_\text{s}/r_{\text{sp}})^{-\gamma _\text{sp} } \rho _\text{sp}}{\gamma _\text{sp} (\gamma _\text{sp} -1)(\gamma _\text{sp} -2)(\gamma _\text{sp}-3)}\right)/r\right]^{1/2}.
\label{e4}
\end{aligned}
\end{equation}
%%%%%%%%%%%%%%%%%%%%%%%%
On the other hand, for a pure dark matter spacetime (the spacetime full of the dark matter), the geometry of that can usually be written as the form of the static spherical symmetric metric,
\begin{equation}
\begin{aligned}
ds^{2}=-f(r)dt^{2}+ g(r)^{-1}dr^{2}+r^{2}(d\theta ^{2}+\sin^{2}\theta d\phi ^{2}),
\label{e5}
\end{aligned}
\end{equation}
where, $f(r)$ means the redshift function and $g(r)$ means the shape function. The relationship between the redshift function $f(r)$ and the tangential velocity $V$ in Ref. \cite{Matos:2003nb} is given by
\begin{equation}
V(r)^2=\frac{r}{\sqrt{f(r)}}\frac{d\sqrt{f(r)}}{dr}=r\frac{dln\sqrt{f(r)}}{dr}
\label{e6}
\end{equation}
Here, we mainly consider the simpler case in Eq. (\ref{e5}), that is $f(r)=g(r)$. According to Eqs. (\ref{e4}) and (\ref{e6}), we can get an analytical solution of the metric coefficient $f(r)$ in a dark matter halo
\begin{equation}
\begin{aligned}
f(r)&=g(r)=\exp \left[-\frac{48 \pi  \rho_\text{sp} r_\text{sp}^{\gamma _{\rm{sp}} } (k R_\text{s})^{3-\gamma _{\rm{sp}} }}{(\gamma _{\rm{sp}} -3) (\gamma _{\rm{sp}} -2) (\gamma _{\rm{sp}} -1) \gamma _{\rm{sp}}  } r^{-1}-\frac{8 \pi  \rho_\text{sp} (k R_\text{s})^3  r_\text{sp}^{\gamma _{\rm{sp}} }}{\gamma _{\rm{sp}}  (\gamma _{\rm{sp}} +1)} r^{-\gamma _{\rm{sp}} -1} \right.\\
&  +\frac{24 \pi  \rho_\text{sp} (k R_\text{s})^2 r_\text{sp}^{\gamma _{\rm{sp}}}}{(\gamma _{\rm{sp}} -1) \gamma _{\rm{sp}} } r^{-\gamma _{\rm{sp}} }        - \frac{24 \pi \rho_\text{sp} k R_\text{s}  r_\text{sp}^{\gamma _{\rm{sp}} }}{(\gamma _{\rm{sp}} -2) (\gamma _{\rm{sp}} -1)} r^{1-\gamma _{\rm{sp}} }+ \left.     \frac{8 \pi  \rho_\text{sp}  r_\text{sp}^{\gamma _{\rm{sp}} }}{(\gamma _{\rm{sp}} -3) (\gamma _{\rm{sp}} -2)} r^{2-\gamma _{\rm{sp}} }        \right].
\label{e7}
\end{aligned}
\end{equation}
%%%%%%%%%%%%%%%%%%%%
For the $f(r)$ in Eq. (\ref{e7}), it is not difficult to find that
\begin{equation}
\lim_{\rho_\text{sp}\rightarrow 0}f(r)  = 1,\lim_{r\rightarrow \infty }f(r) = 1,
\label{e8}
\end{equation}
The Eqs. (\ref{e8}) show that when the density of dark matter is zero or the distance is infinity away from the black hole, the dark matter is absent. Therefore, the spacetime can be approximated as Minkowshi flat spacetime.

\section{The geometry of black holes in a dark matter spike}\label{s3}
In this section, we extend our results (Eq. (\ref{e5})) to the case of black hole in a dark matter spike.  With the methods recorded in Refs. \cite{Xu_2018,Xu:2021dkv,Jusufi:2019nrn}, these authors show that, in Einstein field equations, the energy-momentum tensor of a black hole in a dark matter spike can usually be considered as a linear combination of a vacuum black hole with pure dark matter spacetime. Therefore, we first need to solve the energy-momentum tensor of pure dark matter spacetime, and it satisfy the Einstein field equations
\begin{equation}
R_{\mu \nu }-\frac{1}{2}g_{\mu \nu }R=\kappa  T_{\mu \nu }(\text{DM-spike}),
\label{e9}
\end{equation}
where, $\kappa=8\pi$, $R_{\mu \nu }$ is Ricci tensor and $R$ is Ricci scalar. $T_{\mu\nu}$ is the energy-momentum tensor of dark matter spike and it can be defined as ${T^\nu}_\mu=g^{\nu \alpha }T_{\mu \alpha }=diag[-\rho ,p_r, p, p]$. Taking the metric (\ref{e5}) into Eq. (\ref{e9}), we can obtain
\begin{equation}
\begin{aligned}
&\kappa  {T^{t}}_{t}(\text{DM-spike})=g(r)\left ( \frac{1}{r}\frac{g'(r)}{g(r)}+\frac{1}{r^2} \right ) - \frac{1}{r^2}, \\
&\kappa  {T^{r}}_{r}(\text{DM-spike})=g(r)\left ( \frac{1}{r^2} + \frac{1}{r}\frac{f'(r)}{f(r)} \right )-\frac{1}{r^2},\\
&\kappa  {T^{\theta }}_{\theta }(\text{DM-spike})=\kappa  {T^{\phi }}_{\phi }(\text{DM-spike})\\
&=\frac{1}{2}g(r)\left ( \frac{f''(r)f(r)-f'(r)^2}{f(r)^2}+\frac{f'(r)^2}{2f(r)^2} +\frac{1}{r}(\frac{f'(r)}{f(r)}+\frac{g'(r)}{g(r)}) + \frac{f'(r)g'(r)}{2f(r)g(r)}\right ).
\label{e10}
\end{aligned}
\end{equation}
If considering the black hole in a dark matter spike, the Einstein field equations can be rewritten as
\begin{equation}
R_{\mu \nu }-\frac{1}{2}g_{\mu \nu }R=\kappa (T_{\mu \nu }(\text{BH})+T_{\mu \nu }(\text{DM-spike})).
\label{e11}
\end{equation}
\noindent where, $R_{\mu \nu }$ is Ricci tensor, $g_{\mu \nu }$ is the metric of black holes in a dark matter spike and $R$ is Ricci scalar. Similarly, for a black hole in a dark matter spike, its geometric metric can be written as a linear combination of black hole and dark matter, 
\begin{equation}
\begin{aligned}
ds^{2}=-(f(r)+F1(r))dt^{2}+\frac{1}{g(r)+G1(r)}dr^{2}+r^2(d\theta ^{2}+\sin^{2}\theta d\phi ^{2}),
\label{e12}
\end{aligned}
\end{equation}
where, $f(r)$ and $g(r)$ are the metric coefficients of the pure dark matter spike. Now, taking Eq. (\ref{e12}) into Eq. (\ref{e11}) and using $T_{\mu \nu }(\text{BH})=0$ for a vacuum black hole, we can always obtain the component of the Einstein field equations
\begin{equation}
\begin{aligned}
&(g(r)+G1(r))\left ( \frac{1}{r}\frac{g'(r)+G1'(r)}{g(r)+G1(r)}+\frac{1}{r^2} \right ) =g(r)\left (\frac{1}{r}\frac{g'(r)}{g(r)}+\frac{1}{r^2} \right ),\\
&(g(r)+G1(r))\left ( \frac{1}{r}\frac{f'(r)+F1'(r)}{f(r)+F1(r)}+ \frac{1}{r^2} \right )=g(r)\left ( \frac{1}{r}\frac{f'(r)}{f(r)}+\frac{1}{r^2} \right ).
\label{e13}
\end{aligned}
\end{equation}
Eq. (\ref{e13}) can be described by the combination of two different Einstein field equations components in Eqs. (\ref{e9}) and (\ref{e11}). In the first equation of Eq. (\ref{e13}), only the function $G1(r)$  is unknown. The second equation is based on the functions $G1(r), F1(r)$ and it can be simplified as
\begin{equation}
\begin{aligned}
\frac{f'(r)+F1'(r)}{f(r)+F1(r)} = \frac{g(r)}{g(r)+G1(r)}\left ( \frac{f'(r)}{f(r)}+ \frac{1}{r} \right ) - \frac{1}{r},
\label{e14}
\end{aligned}
\end{equation}
The general solution for the first equation of Eq. (\ref{e13}) is $G1(r)=C/r$, where, $C$ is an undetermined value, which can be determined by boundary condition. Based on Eq.(\ref{e8}), the impacts of dark matter on spacetime at infinity is similar to the case of the Schwarzschild black hole. Therefore, we use the Schwarzschild black hole as the boundary condition at infinity. In other words, the new metric $G(r)=g(r)+G1(r)$ of the black hole in a dark matter spike will degenerate into the Schwarzschild black hole at its boundary ($G(r)=1-2M_\text{BH} /r$). Therefore, a particular solution that satisfies the boundary conditions is $g(r)=1$, $G1(r)=-2M_\text{BH}/r$. Finally, we can obtain the constant value $C=-2M_\text{BH}$. Therefore, the analytical expression of functions $F1(r), G1(r)$ is given by
\begin{equation}
\begin{aligned}
&G1(r)= -\frac{2M_\text{BH}}{r},\\
&F1(r)=\exp\left ( \int \frac{g(r)}{g(r)+G1(r)}\left ( \frac{f'(r)}{f(r)}+ \frac{1}{r}  \right )dr- \frac{1}{r}dr \right )-f(r).
\label{e15}
\end{aligned}
\end{equation}
Using the previous assumption $f(r)=g(r)$ in Eq. (\ref{e13}), we find that $F1(r)=G1(r)= - 2M_\text{BH} / r$. Therefore, the new metric for black holes in a dark matter spike is given by
\begin{equation}
\begin{aligned}
ds^{2}=-F(r)dt^{2}+\frac{1}{G(r)}dr^{2}+H(r)(d\theta ^{2}+\sin^{2}\theta d\phi ^{2}),
\label{e16}
\end{aligned}
\end{equation}
where, $F(r)=f(r)+F1(r), G(r)=g(r)+G1(r), H(r)=r^2$. Finally, the metric coefficients for a black hole in a dark matter spike is given by
\begin{equation}
\begin{aligned}
F(r)&=G(r)=\exp \left[-\frac{48 \pi  \rho_\text{sp} r_\text{sp}^{\gamma _{\rm{sp}} } (k R_\text{s})^{3-\gamma _{\rm{sp}} }}{(\gamma _{\rm{sp}} -3) (\gamma _{\rm{sp}} -2) (\gamma _{\rm{sp}} -1) \gamma _{\rm{sp}}  } r^{-1}-\frac{8 \pi  \rho_\text{sp} (k R_\text{s})^3  r_\text{sp}^{\gamma _{\rm{sp}} }}{\gamma _{\rm{sp}}  (\gamma _{\rm{sp}} +1)} r^{-\gamma _{\rm{sp}} -1} \right.\\
&  +\frac{24 \pi  \rho_\text{sp} (k R_\text{s})^2 r_\text{sp}^{\gamma _{\rm{sp}}}}{(\gamma _{\rm{sp}} -1) \gamma _{\rm{sp}} } r^{-\gamma _{\rm{sp}} }        - \frac{24 \pi \rho_\text{sp} k R_\text{s}  r_\text{sp}^{\gamma _{\rm{sp}} }}{(\gamma _{\rm{sp}} -2) (\gamma _{\rm{sp}} -1)} r^{1-\gamma _{\rm{sp}} }+ \left.     \frac{8 \pi  \rho_\text{sp}  r_\text{sp}^{\gamma _{\rm{sp}} }}{(\gamma _{\rm{sp}} -3) (\gamma _{\rm{sp}} -2)} r^{2-\gamma _{\rm{sp}} }        \right]\\
 &- \frac{2M_\text{BH}}{r},
\label{e17}
\end{aligned}
\end{equation}
If the dark matter is absent, that is $\rho_\mathrm{sp}=0$, we find that the Eq. (\ref{e17}) degenerates into the Schwarzschild black hole, that is $F(r)=G(r)=1-2M_\text{BH}/r$. 

\section{Master wave equations of black holes in a dark matter spike }\label{s4}
\subsection{Scalar field perturbation}
In this subsection, we consider scalar field perturbation of black holes in a dark matter spike. In massless scalar field, for the scalar particles near the black hole, the general covariant equations always satisfy the Klein-Gordon equations
\begin{equation}
\frac{1}{\sqrt{-g}}{\partial_\sigma}(\sqrt{-g}g^{\sigma\nu} {\partial _\nu}\Psi)=0,
\label{e18}
\end{equation}
where, $g^{\sigma\nu}$ is the inverse metric. Taking Eq. (\ref{e16}) into Eq. (\ref{e18}) and using the ansatz
\begin{equation}
\begin{aligned}
\Psi (t,r,\theta ,\phi )=\sum_{l, m} \psi (t,r)Y_{lm}(\theta ,\phi )/r,
\label{e19}
\end{aligned}
\end{equation}
Finally, we can obtain the master wave equations
\begin{equation}
\begin{aligned}
\frac{\partial ^2}{\partial t^2}\psi (t, r)-\frac{\partial ^2}{\partial r_{*}^{2}}\psi (t, r)+V(r)\psi (t, r)=0,
\label{e20}
\end{aligned}
\end{equation}
where, $r_*$ is the tortoise coordinate
\begin{equation}
\begin{aligned}
dr_*=\frac{1}{\sqrt{F(r)G(r)}}dr,
\label{e21}
\end{aligned}
\end{equation}
and the effective potential $V(r)$ reads
\begin{equation}
\begin{aligned}
V(r)=F(r)\frac{l(l+1)}{r^2}+\frac{1}{2r}\frac{\text{d}}{\text{dr}}(F(r)G(r)).
\label{e22}
\end{aligned}
\end{equation}

\subsection{Axial gravitational perturbation}

On the other hand, the perturbations of the black hole can also be considered to be excited by other spin fields, such as gravitational perturbation, electromagnetic field perturbation and neutrino field perturbation. For gravitational perturbation, also called metric perturbation, its metric can be composed of two parts: a background metric and small linear perturbation. The former describes a stable spacetime, while the latter is usually due to the injection of gravitational waves or particles falling into the black hole \cite{Davis:1972ud,1988sfbh.book.....F}. Therefore, for the metric of gravitational perturbations, we have
\begin{eqnarray}
g_{\mu \nu }=\bar{g}_{\mu \nu }+h_{\mu \nu },
\label{e421}
\end{eqnarray}
where, $\bar{g}_{\mu \nu }$ is a background metric of black holes in a dark matter spike in Eq. (\ref{e16}) and $h_{\mu \nu }$ is small linear perturbation ($h_{\mu \nu } \ll \bar{g}_{\mu \nu }$) which reads
\begin{equation}
h_{\mu\nu}=\sum_{l=0}^{\infty }\sum_{m=-l}^{m=l}[(h_{\mu\nu}^{lm})^{\text{axial}}+(h_{\mu\nu}^{lm})^{\text{polar}}],
\label{e422}
\end{equation}
here, we mainly consider axial gravitational perturbation and small linear perturbation $(h_{\mu\nu}^{lm})^{\text{axial}}$ can be written as follows \cite{Regge:1957td}
\begin{eqnarray}
(h_{\mu\nu}^{lm})^{\text{axial}}=\begin{pmatrix}
0 &  0& 0 & h_{0}(t,r)\\
 0& 0 &  0& h_{1}(t,r)\\
0 & 0 & 0 &0 \\
h_{0}(t,r) &h_{1}(t,r)  &0  & 0
\end{pmatrix}{\rm sin}\theta \partial \theta P_{l}({\rm cos}\theta )
\label{e423}
\end{eqnarray}
where, $P_{l}({\rm cos}\theta )$ are the Legendre polynomials of the order $l$ and $m=0$ is the case of the  spherical symmetry. Here, the master wave equation of the axial gravitational perturbation can be given by Einstein field equations
\begin{equation}
\begin{aligned}
E_{\mu\nu} \equiv R_{\mu\nu}-\frac{1}{2} g_{\mu\nu}R=\kappa T_{\mu\nu},
\label{e424}
\end{aligned}
\end{equation}
where, $R_{\mu \nu }$ is Ricci tensor, $R$ is Ricci scalar and $T_{\mu\nu}$ is the energy-momentum tensor of dark matter. Inspired by Refs. \cite{Zhang:2021bdr,Zhao:2023tyo,Zhao:2023itk}, neglecting the perturbation of dark matter, the nonzero components $E_{24}$ and  $E_{34}$ can be written as
\begin{equation}
\begin{aligned}
&\frac{\partial^2}{\partial t ^2}h_1 -\frac{\partial}{\partial r}\frac{\partial}{\partial t}h_0 +\frac{H'}{H} \frac{\partial}{\partial t}h_0 - \frac{G F' H' - F [2 l (l+1) - G' H' - 2 G H'']}{2 H}h_1=0,
\label{e425}
\end{aligned}
\end{equation}
and
\begin{equation}
\begin{aligned}
&-\frac{1}{F}\frac{\partial }{\partial t} h_0 +  G \frac{\partial}{\partial r}h_1 + \frac{F G' + G F'}{2 F}h_1=0,
\label{e426}
\end{aligned}
\end{equation}
where, the functions $h_0, h_1$ are the function of $t$ and $r$, while the metric functions $F, G, H$ are the function of $r$. The prime symbols $'$ and $''$ above equations denote the first and second derivative with respects to $r$. Solving the term $\frac{\partial }{\partial t} h_0$ in Eq. (\ref{e426}) and taking it into Eq. (\ref{e425}), then using the following ansatz
\begin{equation}
\begin{aligned}
\psi (t,r)=\sqrt{F(r)G(r)/H(r)}h_1(t,r),
\label{e427}
\end{aligned}
\end{equation}
the Eqs. (\ref{e425}) and  (\ref{e426}) can be written as a new form
\begin{equation}
\begin{aligned}
\frac{\partial^2 }{\partial t^2} \psi(t,r) - FG\frac{\partial^2 }{\partial r^2} \psi(t,r) -\frac{GF'+FG'}{2} \psi(t,r) + V_\text{eff}(r) \psi(t,r)=0.
\label{e428}
\end{aligned}
\end{equation}
Here, we use tortoise coordinates in Eq. (\ref{e428}), because it can expand the event horizon to infinity. The tortoise coordinates can be given by
\begin{equation}
\begin{aligned}
dr_*=\frac{1}{\sqrt{F(r)G(r)}}dr.
\label{e429}
\end{aligned}
\end{equation}
Finally, we can obtain the master wave equation of the axial gravitational perturbation in a dark matter spike from Eq. (\ref{e428})
\begin{equation}
\begin{aligned}
&\frac{\partial^2 }{\partial t^2} \psi(t,r) - \frac{\partial^2 }{\partial r_*^2} \psi(t,r) + V_\text{eff} (r) \psi(t,r)=0,
\label{e430}
\end{aligned}
\end{equation}
where,
\begin{equation}
\begin{aligned}
V_\text{eff} (r)=\frac{3FG(H')^2}{4H^2} &- \frac{GF'H' +FG'H'+2FGH''}{4H} \\
&- \frac{G F' H'-F [2 l (l+1)-G' H' -2G H'']}{2 H}.
\label{e431}
\end{aligned}
\end{equation}
The functions $F(r), G(r), H(r)$ can be obtained in Eq. (\ref{e16}) and (\ref{e17}). It is particularly noticing that if dark matter is absent ($\rho_\text{sp}=0$), the effective potential in Eq. (\ref{e431}) degenerates into the case of Schwarzschild black hole, that is
\begin{equation}
\begin{aligned}
V_{\text{Sch}}(r)=(1-\frac{2M}{r})(\frac{l(l+1)}{r^2}-\frac{6M}{r^3}).
\label{e432}
\end{aligned}
\end{equation}

\section{The methods for quasinormal mode of black holes in a dark matter spike}\label{s5}
The quasinormal mode (QNM) of black holes in a dark matter spike can be represented by a complex frequency. To obtain QNM, we use the time domain integration method and the WKB method. In the time domain integration method, using the light-cone coordinates $u=t-r_*$ and $v=t+r_*$, the master wave equations of the scalar field perturbation and the axial gravitational perturbation in Eqs. (\ref{e20}) and (\ref{e430}) can be written as 
\begin{equation}
4\frac{\partial ^{2}\psi (u ,v )}{\partial u \partial v } + V_\text{eff}(u ,v)\psi (u ,v )=0,
\label{e21}
\end{equation}
where, $r_*$ in the light-cone coordinates denotes the tortoise coordinates. For Eq. (\ref{e21}), a reasonable discretization scheme in Refs. \cite {PhysRevD.63.084014,PhysRevD.64.044024,PhysRevD.72.044027} is given by
\begin{equation}
\Psi(N)=\Psi(W)+\Psi(E)-\Psi(S)-h ^2\frac{V(W)\psi(W)+V(E)\psi(E)}{8}+O(h ^4),     
\label{e22}                                             
\end{equation}
where, $h$ is step size of each grid cell and the integration grid are $ N=(u+h, v+h)$, $W=(u+h, v)$, $E=(u, v +h)$ and $S = (u, v)$, respectively. Then, the Gaussian pulse is applied on two null surfaces, $u = u_0$ and $v = v_0$,
\begin{equation}
\begin{aligned}
\psi (u=u _{0} ,v )=A {\rm exp}\left [ -\frac{(v -v _{0})^{2}}{\sigma ^{2}} \right ], \quad \psi (u ,v=v_{0} )=0,  
\label{e23}     
\end{aligned} 
\end{equation}
where, $A=1$, $v_0=10$ and $\sigma=3$. In this way, the time evolution of a black hole under the scalar perturbation can be obtained. Finally, in the time evolution, ones can use the Prony method to extract the QNM frequencies \cite{RevModPhys.83.793,PhysRevD.75.124017,Chowdhury:2020rfj} using
\begin{eqnarray}
\psi(t)\simeq \sum ^p_{i=1}C_ie^{-i\omega_it}.                  
\label{e24}                    
\end{eqnarray}

On the other hand, when calculating QNM frequencies, the WKB method is also a widely used method. This method was first proposed by Schutz and Will \cite{Schutz:1985km}, and then promoted by Iyer, Will, and Konoplya \cite{PhysRevD.35.3621,PhysRevD.35.3632,PhysRevD.68.024018,PhysRevD.68.124017,RevModPhys.83.793}. In this work, we use the sixth order WKB formula in Ref. \cite{PhysRevD.68.124017,PhysRevD.68.024018}
\begin{eqnarray}
\frac{i (\omega ^{2}-V_{0})}{\sqrt{-2V_{0}^{''}}}-\sum ^{6}_{i=2}\Lambda _{i}=n+\frac{1}{2}, \quad(n=0,1,2, \cdots)    
\label{e25}           
\end{eqnarray}
where, $V_{0}$ is the maximum of the effective potential, the prime is the second derivative of the effective potential $V$ at its maximum, $\Lambda _{i}$ is the $i$th order revision terms which can be obtained in Ref. \cite{PhysRevD.68.024018} and $n$ is the overtone number. In this work, we mainly calculate the frequencies of fundamental quasinormal mode, that is the case of $n=0$. 

\begin{table}[t!]
\setlength{\abovecaptionskip}{0.2cm}
\setlength{\belowcaptionskip}{0.2cm}
\centering
\caption{The fundamental QNM of Schwarzschild black hole in the scalar field and axial gravitational perturbation using Prony method, the sixth order WKB method and the Leaver's results in Ref. \cite{Iyer:1986nq}.}
\scalebox{.96}{
\begin{tabular}{cccc}
\hline \hline
      modes $(M_{\text{BH}}=1/2) $                       &  Leaver's results              & Prony method                   & WKB method                 \\
 \hline
$n=0, l_{\text{scalar}} =0$                    & 0.2210 - 0.2098 i & 0.232173 - 0.216358 i & 0.219199 - 0.219982 i \\
$n=0, l_{\text{scalar}} =1$                    & 0.5858 - 0.1954 i & 0.580419 - 0.202386 i & 0.585907 - 0.194772 i \\
$n=0, l_{\text{scalar}} =2$                    & 0.9672 - 0.1936 i & 0.953006 - 0.192696 i & 0.967295 - 0.193435 i \\
$n=0, l_{\text{axial}} =2$                      & 0.7474 - 0.1780 i & 0.746541 - 0.176440 i & 0.747239 - 0.177782 i \\
$n=0, l_{\text{axial}} =3$                      & 1.1988 - 0.1854 i & 1.200819 - 0.183197 i & 1.198890 - 0.185405 i \\
\hline \hline
\end{tabular}}
\label{ttt1}
\end{table}

\section{The numerical results}\label{s6}
\subsection{The quasinormal modes of black holes in a DM spike of M87 }\label{s61}
In this subsection, we study the quasinormal mode (QNM) of black holes in a dark matter spike of M87 under the scalar field perturbation and axial gravitational perturbation, then compare them with Schwarzschild black hole. Before that, in order to ensure the accuracy of the results, we first calculate the quasinormal frequencies of Schwarzschild black hole using the sixth order WKB method and the Prony method, which can be found in Table \ref{ttt1}. According to the quasinormal frequencies of Schwarzschild black hole given in Tables I and III in Ref. \cite{Iyer:1986nq}, under the same angular quantum number, for the Prony method, the difference between our results and Leaver's results in the real part of the QNM is approximately from $0.2\%$  to $1\%$ and the imaginary part is approximately from $0.1\%$ to $0.7\%$. Similarly, for the sixth order WKB method, the difference between them in the real part of the QNM is approximately from  $0.01\%$  to $0.2\%$ and the imaginary part is approximately from $0.01\%$ to $1\%$. These results show that the sixth order WKB method and the Prony method we used are in good agreement with Leaver's results. To some extent, therefore, it is reasonable for us to use these methods to study the QNM of the black holes in a dark matter spike.
\begin{figure*}[t!]
\centering
\subfigure[Schw]{\label{fig:a}
\includegraphics[width=0.31 \columnwidth]{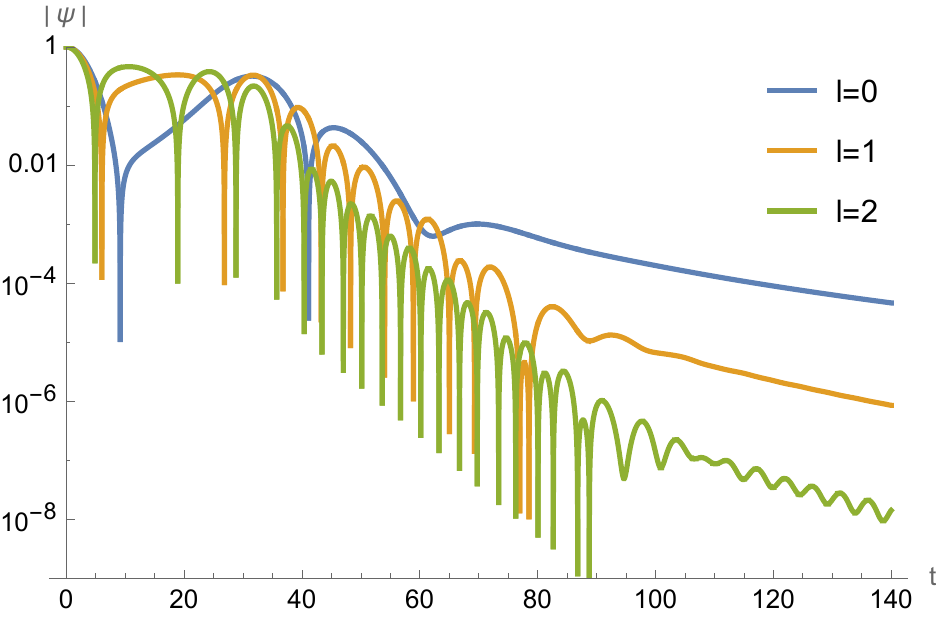}
}
\subfigure[DM, k=2]{\label{fig:b}
\includegraphics[width=0.31 \columnwidth]{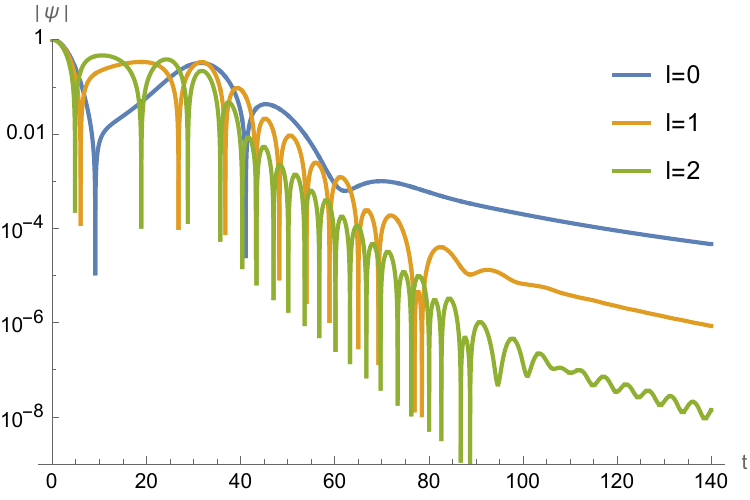}
}
\subfigure[DM, k=4]{\label{fig:c}
\includegraphics[width=0.31 \columnwidth]{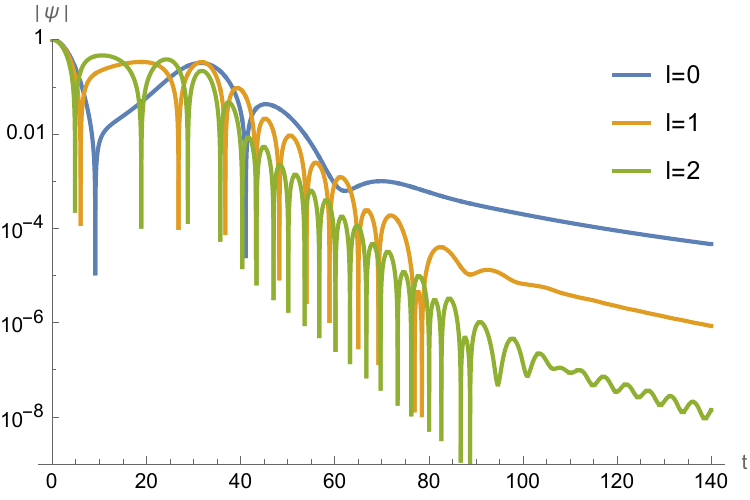}
}
\subfigure[Schw]{\label{fig:a}
\includegraphics[width=0.31 \columnwidth]{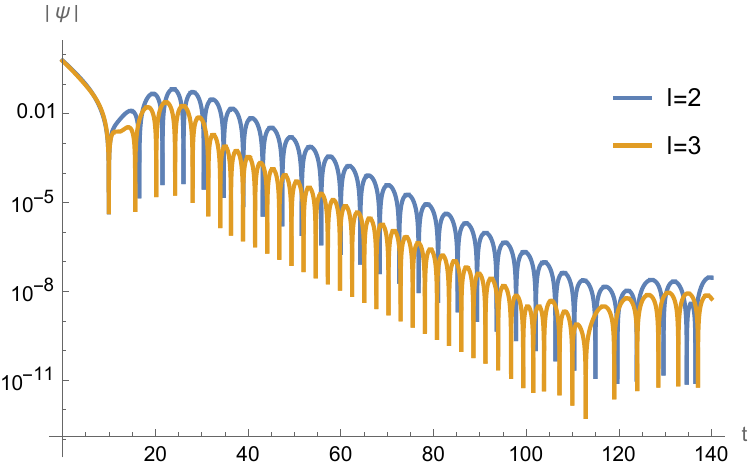}
}
\subfigure[DM, k=2]{\label{fig:b}
\includegraphics[width=0.31 \columnwidth]{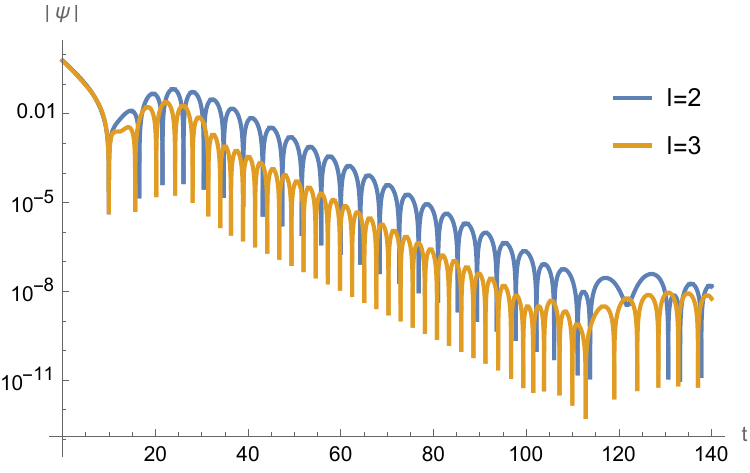}
}
\subfigure[DM, k=4]{\label{fig:c}
\includegraphics[width=0.31 \columnwidth]{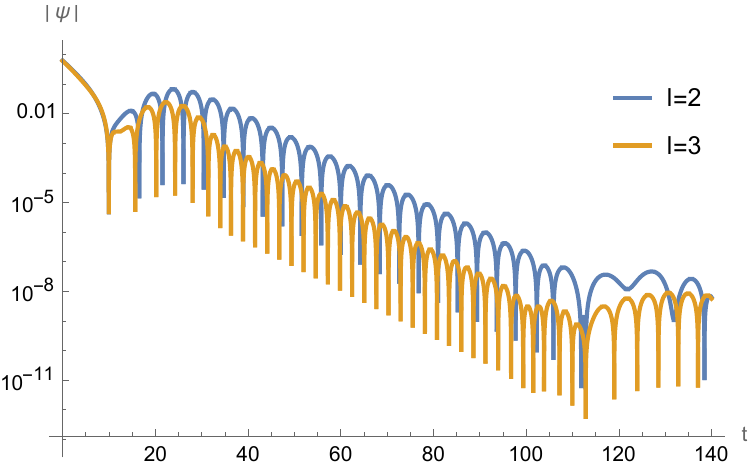}
}
\caption{The time evolution of quasinormal modes in the scalar field perturbation (top panels) and axial gravitational perturbation (bottom panels) with different angular quantum number at the galactic center of M87. The parameters we used are $M_\text{BH}=1/2, \gamma=1/2, \alpha_\gamma=0.1, \rho_0=6.9 \times 10^6 M_\odot/\text{kpc}^3, r_0=91.2\text{kpc}$.}
\label{f2}
\end{figure*}

Now, back to our research. In Figure \ref{f2}, we present the time evolution of the QNM of black holes at the galactic center of M87 with different angular quantum numbers both in the case of a DM spike and the vacuum, respectively. The time evolutions of these QNMs mainly follows three stages, that is initial excitation, QNM, and power-law tail. Here, we focus on the stage of QNM, because this stage can be determined by the black hole's parameters. Besides, in the scalar field perturbation, with the increasing of the angular quantum number, the duration of the QNM stage obviously increases. However, from these figures, it is difficult to distinguish the impacts of dark matter on the QNM of black holes. Therefore, we use the Prony method to extract the QNM frequency of the QNM stage, and record the results in Table \ref{t1} and \ref{nt2}. Our results show that both in the scalar field and axial gravitational perturbation, because of the existence of dark matter, the QNM of black holes is generally smaller than that of the vacuum. In other words, dark matter does affect the QNM of black holes. For these two different dark matter spikes, the numerical results of Newtonian approximation ($k=4$) are greater than the full relativity ($k=2$). With the development of the detection technology, these research results may provide some help in establishing the final dark matter model.

\begin{table}[t!]
\setlength{\abovecaptionskip}{0.2cm}
\setlength{\belowcaptionskip}{0.2cm}
\centering
\caption{The fundamental QNM from the time evolution of black holes in the scalar field perturbation (Figure. \ref{f2}) at the galactic center of M87 using the Prony method.}
\scalebox{.95}{
\begin{tabular}{cccc}
\hline \hline
Types of BHs & $l=0$                 & $l=1$                 & $l=2$                 \\
 \hline
Schw                             & 0.232173 - 0.216358 i & 0.580419 - 0.202386 i & 0.953006 - 0.192696 i \\
DM, $k=2$                    & 0.232172 - 0.216359 i & 0.580416 - 0.202386 i & 0.953001 - 0.192695 i \\
DM, $k=4$                    & 0.232175 - 0.216360 i & 0.580417 - 0.202387 i & 0.953002 - 0.192696 i \\
\hline \hline
\end{tabular}}
\label{t1}
\end{table}
\begin{table}[t!]
\setlength{\abovecaptionskip}{0.2cm}
\setlength{\belowcaptionskip}{0.1cm}
\centering
\caption{The fundamental QNM from the time evolution of black holes in the axial gravitational perturbation (Figure. \ref{f2}) at the galactic center of M87 using the Prony method.}
\scalebox{1}{
\begin{tabular}{ccc}
\hline \hline
Types of BHs                               & $l=2$                          & $l=3$                                \\
 \hline
Schw                             & 0.7465416 - 0.1764407 i & 1.2008195 - 0.18319747 i  \\
DM, $k=2$                    & 0.7465377 - 0.1764411 i & 1.2008133 - 0.18319763 i  \\
DM, $k=4$                    & 0.7465378 - 0.1764418 i & 1.2008139 - 0.18319791 i  \\
\hline \hline
\end{tabular}}
\label{nt2}
\end{table}
On the other hand, we also study the impacts of some DM parameters on the QNM of black holes and compared them with the Schwarzschild black hole. The QNM of black holes in DM spike and the Schwarzschild black hole are calculated by the WKB method. Firstly, one way we test the accuracy of the WKB method is to compare it with the Prony method since it is a semi-analytical method. From the results in Tables. \ref{t2} - \ref{gt7}, under the same parameter of the Schwarzschild black hole,  the difference between the WKB method and the Prony method in the real part of the QNM is approximately from  $0.1\%$  to $1\%$ and the imaginary part is approximately from $0.1\%$ to $0.8\%$, which shows that the WKB method and the Prony method are still in good agreement.

Secondly, in the scalar field and axial gravitational perturbation, we use these two methods to test the impacts of the density parameters, power law index under the different models of DM spike on the QNM of black holes, respectively. For the scalar field perturbation in Tables \ref{t2} - \ref{t4}, when the angular quantum numbers $l=0$, under the case of the the full relativity $k=2$, if the density parameter is fixed, both the QNM frequency and decay rate of black holes in DM spike decrease with the increasing of the power-law index. If the power-law index is fixed, both the QNM frequency and the decay rate of black holes in DM spike decrease with the increasing of the density parameter. Under the Newtonian approximation $k=4$, if the density parameter is fixed, the QNM frequency of black holes in DM spike decreases with the increasing of the power-law index, but the decay rate increases. If the power-law index is fixed, the QNM frequency of black holes in DM spike decreases with increasing of the density parameter, but the decay rate increases. When the angular quantum numbers $l=1$, under the case of the the full relativity $k=2$, if the density parameter is fixed, the QNM frequency and decay rate of black holes in DM spike decrease with the increasing of the power-law index. If the power-law index is fixed, both the QNM frequency and the decay rate of black holes in DM spike decrease with the increasing of the density parameter. Under the Newtonian approximation $k=4$, the variation trend in the black hole QNM are basically the same as $k=2$. The variation trend of QNM of the black hole with angular quantum number $l=2$ is basically the same as that of black hole with angular quantum number $l=1$. 
\begin{table}[t!]
\setlength{\abovecaptionskip}{0.2cm}
\setlength{\belowcaptionskip}{0.2cm}
\centering
\caption{The fundamental QNM ($n=0, l=0$) of the scalar field perturbation as a function of the halo density $\rho_0 (10^6M_\odot/\text{kpc}^3)$ at the fixed radius $r_0=91.2 \text{kpc}$ calculated by the 6th order WKB method. The corresponding QNM of Schwarzschild black hole is $\omega_\text{Sch} =0.219199 - 0.219982 i, M_\text{BH}=1/2$.}
\scalebox{.58}{
\begin{tabular}{ccccccc}
\hline \hline 
  & \multicolumn{3}{c}{$k=2$}                                                                                                          & \multicolumn{3}{c}{$k=4$}                                                                                                            \\
\hline
$\rho_0$ & $\gamma=1/2$                               & $\gamma=2/2$                               & $\gamma=3/2$                               & $\gamma=1/2$                               & $\gamma=2/2$                               & $\gamma=3/2$                               \\
\hline 
6.9                  & 0.219197 - 0.219981 i & 0.219187 - 0.219972 i & 0.219067 - 0.219877 i & 0.219197 - 0.2199827 i & 0.219186 - 0.219989 i & 0.219059 - 0.220137 i \\
100                 & 0.219195 - 0.219979 i & 0.219169 - 0.219957 i & 0.218816 - 0.219676 i & 0.219195 - 0.2199831 i & 0.219169 - 0.219999 i & 0.218793 - 0.220418 i  \\
500                 & 0.219193 - 0.219977 i & 0.219148 - 0.219939 i & 0.218470 - 0.219399 i & 0.219193 - 0.2199835 i & 0.219147 - 0.220011 i & 0.218430 - 0.220777 i  \\
1000               & 0.219192 - 0.219976 i & 0.219135 - 0.219927 i & 0.218238 - 0.219213 i & 0.219192 - 0.2199838 i & 0.219134 - 0.220018 i & 0.218188 - 0.220999 i  \\
\hline \hline 
\end{tabular}}
\label{t2}
\end{table}

\begin{table}[t!]
\setlength{\abovecaptionskip}{0.2cm}
\setlength{\belowcaptionskip}{0.2cm}
\centering
\caption{The fundamental QNM ($n=0, l=1$) of the scalar field perturbation as a function of the halo density $\rho_0 (10^6M_\odot/\text{kpc}^3)$ at the fixed radius $r_0=91.2 \text{kpc}$ calculated by the 6th order WKB method. The corresponding QNM of Schwarzschild black hole is $\omega_\text{Sch}  =0.585907 - 0.194772 i, M_\text{BH}=1/2$.}
\scalebox{.58}{
\begin{tabular}{ccccccc}
\hline \hline 
  & \multicolumn{3}{c}{$k=2$}                                                                                                            & \multicolumn{3}{c}{$k=4$}                                                                                                            \\
\hline
$\rho_0$ & $\gamma=1/2$                               & $\gamma=2/2$                               & $\gamma=3/2$                               & $\gamma=1/2$                               & $\gamma=2/2$                               & $\gamma=3/2$                               \\
\hline 
6.9                  & 0.585904 - 0.194771 i & 0.585882 - 0.194761 i & 0.585637 - 0.194660 i & 0.585904 - 0.1947714 i & 0.585884 - 0.194770 i & 0.585656 - 0.194784  i \\
100                 & 0.585900 - 0.194769 i & 0.585846 - 0.194746 i & 0.585120 - 0.194446 i & 0.585900 - 0.1947711 i & 0.585850 - 0.194768 i & 0.585174 - 0.194806  i \\
500                 & 0.585896 - 0.194767 i & 0.585802 - 0.194728 i & 0.584409 - 0.194152 i & 0.585896 - 0.1947770 i & 0.585810 - 0.194765 i & 0.584512 - 0.194833  i \\
1000               & 0.585893 - 0.194766 i & 0.585775 - 0.194717 i & 0.583931 - 0.193954 i & 0.585894 - 0.1947699 i & 0.585785 - 0.194763 i & 0.584067 - 0.194849  i \\
\hline \hline 
\end{tabular}}
\label{t3}
\end{table}

\begin{table}[t!]
\setlength{\abovecaptionskip}{0.2cm}
\setlength{\belowcaptionskip}{0.2cm}
\centering
\caption{The fundamental QNM ($n=0, l=2$) of the scalar field perturbation as a function of the halo density $\rho_0 (10^6M_\odot/\text{kpc}^3)$ at the fixed radius $r_0=91.2 \text{kpc}$ calculated by the 6th order WKB method. The corresponding QNM of Schwarzschild black hole is $\omega_\text{Sch}  =0.967295 - 0.193435 i, M_\text{BH}=1/2$.}
\scalebox{.58}{
\begin{tabular}{ccccccc}
\hline \hline 
  & \multicolumn{3}{c}{$k=2$}                                                                                                            & \multicolumn{3}{c}{$k=4$}                                                                                                            \\
\hline
$\rho_0$         & $\gamma=1/2$                          & $\gamma=2/2$                               & $\gamma=3/2$                               & $\gamma=1/2$                               & $\gamma=2/2$                               & $\gamma=3/2$                               \\
\hline 
6.9                  & 0.967289 - 0.193434 i & 0.967254 - 0.193424 i & 0.966862 - 0.193322 i & 0.967290 - 0.193434 i & 0.967259 - 0.193432 i & 0.966913 - 0.193422 i  \\
100                 & 0.967283 - 0.193432 i & 0.967196 - 0.193409 i & 0.966033 - 0.193107 i & 0.967284 - 0.193434 i & 0.967207 - 0.193427 i & 0.966182 - 0.193397 i  \\
500                 & 0.967276 - 0.193430 i & 0.967126 - 0.193391 i & 0.964895 - 0.192810 i & 0.967278 - 0.193433 i & 0.967145 - 0.193421 i & 0.965179 - 0.193360 i  \\
1000               & 0.967272 - 0.193429 i & 0.967082 - 0.193380 i & 0.964129 - 0.192611 i & 0.967274 - 0.193432 i & 0.967106 - 0.193417 i & 0.964505 - 0.193334 i  \\
\hline \hline 
\end{tabular}}
\label{t4}
\end{table}
For the axial gravitational perturbation in Tables \ref{gt6} - \ref{gt7}, when angular quantum number $l=2$, under the case of the the full relativity $k=2$, if the density parameter is fixed, the QNM frequency and decay rate of black holes in DM spike decrease with the increasing of the power-law index. If the power-law index is fixed, both the QNM frequency and the decay rate of black holes in DM spike decrease with the increasing of the density parameter. Under the Newtonian approximation $k=4$, the variation trend in the black hole QNM are basically the same as $k=2$. The variation trend of QNM of the black hole with angular quantum number $l=3$ is basically the same as that of black hole with angular quantum number $l=2$.

Finally, in the scalar field and axial gravitational perturbation, we compare these two different QNM of black holes in DM spike with the Schwarzschild black hole. We find that the frequency and absolute value of the decay rate of the QNM of Schwarzschild black holes are almost always larger than those of black holes in DM spike. This is basically consistent with a conclusion from the time evolution of black holes. In other words, dark matter does affect the QNM of black holes. In the next subsection,  from space-based detectors, we will study and explore the possibility of space-based detectors detecting the impacts of dark matter on the QNM of black holes.

\begin{table}[t!]
\setlength{\abovecaptionskip}{0.2cm}
\setlength{\belowcaptionskip}{0.2cm}
\centering
\caption{The fundamental QNM ($n=0, l=2$) of the axial gravitational perturbation as a function of the halo density $\rho_0 (10^6M_\odot/\text{kpc}^3)$ at the fixed radius $r_0=91.2 \text{kpc}$ calculated by the 6th order WKB method. The corresponding QNM of Schwarzschild black hole is $\omega_\text{Sch}  =0.747239 - 0.177782 i, M_\text{BH}=1/2$.}
\scalebox{.58}{
\begin{tabular}{ccccccc}
\hline \hline 
  & \multicolumn{3}{c}{$k=2$}                                                                                                            & \multicolumn{3}{c}{$k=4$}                                                                                                            \\
\hline
$\rho_0$         & $\gamma=1/2$                          & $\gamma=2/2$                               & $\gamma=3/2$                               & $\gamma=1/2$                               & $\gamma=2/2$                               & $\gamma=3/2$                               \\
\hline 
6.9                  & 0.747235 - 0.177781 i & 0.747214 - 0.177773 i & 0.746969 - 0.177684 i & 0.747235 - 0.1777816 i & 0.747211 - 0.177781 i & 0.746929 - 0.177805 i  \\
100                 & 0.747231 - 0.177779 i & 0.747177 - 0.177760 i & 0.746454 - 0.177497 i & 0.747231 - 0.1777812 i & 0.747172 - 0.177780 i & 0.746338 - 0.177846 i  \\
500                 & 0.747227 - 0.177778 i & 0.747134 - 0.177744 i & 0.745746 - 0.177240 i & 0.747226 - 0.1777808 i & 0.747125 - 0.177779 i & 0.745528 - 0.177900 i  \\
1000               & 0.747225 - 0.177777 i & 0.747107 - 0.177734 i & 0.745269 - 0.177067 i & 0.747224 - 0.1777805 i & 0.747095 - 0.177778 i & 0.744986 - 0.177934 i  \\
\hline \hline 
\end{tabular}}
\label{gt6}
\end{table}

\begin{table}[t!]
\setlength{\abovecaptionskip}{0.2cm}
\setlength{\belowcaptionskip}{0.2cm}
\centering
\caption{The fundamental QNM ($n=0, l=3$) of the axial gravitational perturbation as a function of the halo density $\rho_0 (10^6M_\odot/\text{kpc}^3)$ at the fixed radius $r_0=91.2 \text{kpc}$ calculated by the 6th order WKB method. The corresponding QNM of Schwarzschild black hole is $\omega_\text{Sch}  =1.19889 - 0.185405 i, M_\text{BH}=1/2$.}
\scalebox{.6}{
\begin{tabular}{ccccccc}
\hline \hline 
  & \multicolumn{3}{c}{$k=2$}                                                                                                            & \multicolumn{3}{c}{$k=4$}                                                                                                            \\
\hline
$\rho_0$         & $\gamma=1/2$                          & $\gamma=2/2$                               & $\gamma=3/2$                               & $\gamma=1/2$                               & $\gamma=2/2$                               & $\gamma=3/2$                               \\
\hline 
6.9                  & 1.198880 - 0.185404 i & 1.19884 - 0.185395 i & 1.19840 - 0.185299 i & 1.198880 - 0.1854044 i & 1.19884 - 0.185402 i & 1.19842 - 0.185399 i  \\
100                 & 1.198873 - 0.185402 i & 1.19878 - 0.185381 i & 1.19747 - 0.185097 i & 1.198874 - 0.1854038 i & 1.19878 - 0.185398 i & 1.19754 - 0.185386 i  \\
500                 & 1.198865 - 0.185400 i & 1.19870 - 0.185364 i & 1.19618 - 0.184819 i & 1.198867 - 0.1854031 i & 1.19871 - 0.185394 i & 1.19632 - 0.185365 i  \\
1000               & 1.198861 - 0.185399 i & 1.19865 - 0.185353 i & 1.19532 - 0.184632 i & 1.198862 - 0.1854027 i & 1.19866 - 0.185391 i & 1.19551 - 0.185350 i  \\
\hline \hline 
\end{tabular}}
\label{gt7}
\end{table}

\subsection{The detectability of the impacts on the QNM in a dark matter spike}\label{s62}
In the above subsection, we studied and analyzed the fundamental quasinormal mode (QNM) of black holes in the DM spike of M87, and studied the impacts of the DM parameters on the QNM frequencies. Therefore, here, we are interested in the possibility of space-based detection detecting the impacts of dark matter on the QNM of black holes. In the axial gravitational perturbation, the ringdown waveform of gravitational waves in Ref. \cite{Berti:2005ys} can be written as
\begin{equation}
h(t)=h_+(t)+ih_\times (t)=\frac{M}{D_r}\sum_{lmn}A_{lmn}e^{i(f_{lmn}t+\phi_{lmn})e^{-t/\tau_{lmn}}}S_{lmn},
\end{equation}
where, $M$ means the redshift mass of black holes, $D_r$ means the luminosity distance to the source, $A_{lmn}$ means the amplitude of the QNM,  $\phi_{lmn}$ means the phase coefficient and $S_{lmn}$ means the spheroidal harmonics of the spin weight $2$. The frequency $f_{lmn}^{\text{DM}}$ and the decay time $\tau_{lmn}^{\text{DM}}$ of the QNM in DM spike are given by
\begin{equation}
2\pi f_{lmn}^{\text{DM}}=\text{Re}(\omega_{lmn}^{\text{DM}}), \quad \tau_{lmn}^{\text{DM}}=-1/\text{Im}(\omega_{lmn}^{\text{DM}}),
\end{equation}
where, $\omega_{lmn}^{\text{DM}}$ means the fundamental QNM of black holes in a DM spike of M87 under the axial gravitational perturbation. In General Relativity, the ringdown waveform of gravitational waves can be obtained starting from quadrupole radiation ($l=2$), where, the modes $l=2$ and $l=3$ are expected to the dominant mode of gravitational waves signal \cite{Berti:2005ys,Berti:2009kk}. Dark matter as an additional source, it causes deviations from the black hole in General Relativity. Based on the general solution method given in Refs. \cite{Zhang:2021bdr,Zhao:2023tyo,Zhao:2023itk}, we ignore the perturbation of dark matter, and then the impacts of dark matter is only reflected in the black hole metric. Therefore, the dominant mode of axial gravitational perturbation in a dark matter spike is similar to the Schwarzschild black hole. Finally, in these two modes $l=2$ and $l=3$, we choose $(l,m,n)=(2,0,0)$ as the dominant mode because this mode has the longest decay time, that is, the imaginary part of the quasinormal mode frequency is the smallest (See Tables \ref{gt6} and \ref{gt7}). Based on the Ref. \cite{Zhang:2022roh}, the frequency and decay time of black holes in a DM spike of M87 is given by
\begin{equation}
f_{lmn}^\text{DM}=f_{lmn}^{\text{Sch}}(1+\delta f_{lmn}),\quad \tau_{lmn}^\text{DM}=\tau_{lmn}^{\text{Sch}}(1+\delta \tau_{lmn}),
\end{equation}
where, $f_{lmn}^{\text{Sch}}, \tau_{lmn}^{\text{Sch}}$ are the frequency and damping time of the QNM in the Schwarzschild black hole. $\delta f_{lmn},\delta \tau_{lmn}$ are the relative deviations to the case of the Schwarzschild black hole. Our results in Tables \ref{t5} and  \ref{t6} show that whether under Newtonian approximation or full relativity, the relative deviations of QNM frequency and damping time increase with the increasing of the power-law index. The relative deviation is approximately $10^{-6} \sim  10^{-4}$ for the frequency and $10^{-6} \sim  10^{-4}$ for the damping time using the Prony method. For the results of the WKB method, the relative deviation is approximately $10^{-6} \sim  10^{-4}$ for the frequency and $10^{-7} \sim  10^{-4}$ for the damping time. From these results, the impacts of dark matter spike on the QNM of black holes can reach up to $10^{-4}$. In Ref. \cite{Shi:2019hqa}, C. Shi et al. showed that the current TianQin observatory can detect the maximum relative deviations of frequency and damping time to $10^{-3}$. This result is very close to our predicted results although there are still some gaps. Different from the dark matter halo without a spike, N. Dai et al. showed that the ringdown of the gravitational waves can produce relative deviations less than $10^{-5}$ between the presence and absence of dark matter \cite{Dai:2023cft}. In the near future, with the continuous development of observation technology, we believe that a new generation of detectors will be able to detect even smaller deviations. We hope that these research results will provide some help for the indirect detection of dark matter.

\begin{table}[t!]
\setlength{\abovecaptionskip}{0.2cm}
\setlength{\belowcaptionskip}{0.2cm}
\centering
\caption{The relative deviations of the frequency $f_{200}$ and damping time $\tau_{200}$ in dark matter spike of M87 with different power-law index $\gamma$ using the Prony method. The main calculation parameters are $M_\text{BH}=1/2, l=2, \alpha_\gamma=0.1$, $\rho_0=6.9 \times 10^{6}$ $M_{\bigodot}/\text{kpc}^{3}$, $r_0=91.2$ $\text{kpc}$.}
\scalebox{.9}{
\begin{tabular}{ccccccccc}
\hline \hline
  & \multicolumn{3}{c}{$k=2$}                                                                                                            & \multicolumn{3}{c}{$k=4$}     \\
 \hline
                                       & $\gamma=1/2 $           & $\gamma=2/2 $               & $\gamma=3/2 $                & $\gamma=1/2  $         & $\gamma=2/2$                 & $\gamma=3/2$              \\ 
   \hline
$\delta f_{200}$       & 5.20 $\times 10^{-6}$ & 3.83 $\times 10^{-5}$      & 4.20 $\times 10^{-4}$    & 5.10 $\times 10^{-6}$    & 3.83 $\times 10^{-5}$    & 4.87 $\times 10^{-4}$       \\
$\delta \tau_{200}$  & 2.49 $\times 10^{-6}$ & 1.87 $\times 10^{-5}$      & 2.36 $\times 10^{-4}$    & 6.14 $\times 10^{-6}$    & 5.39 $\times 10^{-5}$    & 7.97 $\times 10^{-4}$        \\ 
\hline \hline
\end{tabular}}
\label{t5}
\end{table}
\begin{table}[t!]
\setlength{\abovecaptionskip}{0.2cm}
\setlength{\belowcaptionskip}{0.2cm}
\centering
\caption{The relative deviations of the frequency $f_{200}$ and damping time $\tau_{200}$ in dark matter spike of M87 with different power-law index $\gamma$ using the WKB method. The main calculation parameters are $M_\text{BH}=1/2, l=2,  \alpha_\gamma=0.1$, $\rho_0=6.9 \times 10^{6}$ $M_{\bigodot}/\text{kpc}^{3}$, $r_0=91.2$ $\text{kpc}$.}
\scalebox{.9}{
\begin{tabular}{ccccccccc}
\hline \hline
  & \multicolumn{3}{c}{$k=2$}                                                                                                            & \multicolumn{3}{c}{$k=4$}     \\
 \hline
                                    & $\gamma=1/2 $            & $\gamma=2/2 $                & $\gamma=3/2 $               & $\gamma=1/2  $           & $\gamma=2/2$             & $\gamma=3/2$              \\ 
   \hline
$\delta f_{200}$       & 5.54 $\times 10^{-6}$ & 4.08 $\times 10^{-5}$      & 4.37 $\times 10^{-4}$    & 6.15 $\times 10^{-6}$    & 4.74 $\times 10^{-5}$    & 5.46 $\times 10^{-4}$       \\
$\delta \tau_{200}$  & 7.10 $\times 10^{-6}$ & 5.18 $\times 10^{-5}$      & 5.47 $\times 10^{-4}$    & 1.24 $\times 10^{-7}$    & 1.30 $\times 10^{-5}$    & 3.83 $\times 10^{-4}$        \\ 
\hline \hline
\end{tabular}}
\label{t6}
\end{table}

The last but not the least, from the order of magnitude of these relative deviations, DM spike may have the significant impacts on the QNM of black holes compared to density profiles without the DM spike in Refs. \cite{Liu:2022ygf,Liu:2023vno}.

\section{\label{s7} Summary}
In this work, we study the quasinormal mode (QNM) of black holes in a dark matter (DM) spike of M87 under the scalar field and axial gravitational perturbation, and then compare them with Schwarzschild black hole.  Meanwhile, the impacts of DM density parameters on the QNM of black holes have been studied. Finally, we explore the possibility of space-based detectors detecting the impacts of DM spike on the QNM of black holes.  The main summaries are as follows:\\
\indent (1) From Einstein field equations, we construct a set of solutions of the Schwarzschild-like black hole in a DM spike, and these solutions are given by Eq. (\ref{e17}). Combined with the mass model of M87, we numerically calculate the time evolution and the QNM frequency of black holes under the scalar field and axial gravitational perturbation, and then compare them with Schwarzschild black hole. The frequency and decay rate of the QNM of the Schwarzschild black hole are both larger than that of black holes in a DM spike, indicating that dark matter may have the significant impacts on the QNM of black holes. Under the same DM parameters, the frequency and decay rate of the QNM under the full relativity are basically smaller than under the Newtonian approximation.

(2) The impacts of some parameters of DM spike on the QNM of black holes have been studied in the scalar field and axial gravitational perturbation. In particular, when the angular quantum numbers and density parameter are fixed under the full relativity $k=2$, both the QNM frequency and decay rate of black holes in DM spike decrease with the increasing of the power-law index $\gamma$ . When the power-law index is fixed, both the QNM frequency and the decay rate of black holes in DM spike decrease with the increasing of the density parameter. Under the Newtonian approximation $k=4$, the variation trend in the QNM of black holes are basically the same as $k=2$. In the axial gravitational perturbation, the variation trend in the QNM of black holes are basically the same as the scalar field perturbation. These new features from the QNM of black holes under the Newtonian approximation and full relativity may provide some help for the establishment of the final dark matter model to a certain extent.

(3) The possibility of space-based detectors detecting the impact of DM spike on the QNM of black holes have been explored in the axial gravitational perturbation. Whether under Newtonian approximation or full relativity, the relative deviations of QNM frequency and damping time increase with the increasing of the power-law index $\gamma$. The relative deviation is approximately $10^{-6} \sim 10^{-4 }$ for the frequency and $10^{-6} \sim 10^{-4}$ for the damping time using the Prony method. For the WKB method, the relative deviation is approximately $10^{-6} \sim 10^ {-4}$ for the frequency and $10^{-7} \sim 10^{-4}$ for the damping time. From these results, the impacts of dark matter spike on the QNM of black holes can reach up to $10^{-4}$. The current TianQin observatory can detect the maximum relative deviations to $10^{-3}$. This result is very close to our predicted results although there are still some gaps. In the near future, with the continuous development of observation technology, we believe that a new generation of detectors will be able to detect even smaller deviations. These results of black holes in a DM spike may provide some help for the indirect detection of dark matter.

\section*{Acknowledgments}
We would like to acknowledge the anonymous referees for the constructive reports, which significantly improved this paper. This research was partly supported by the National Natural Science Foundation of China (Grants No. 12265007).

%\section*{References}
%\bibliographystyle{model1a-num-names}
\bibliographystyle{elsarticle-num}
\bibliography{plbexample}
%%%%%%%%%%%%%%%%%%%%%%%

\end{document}